# Perches, Post-holes and Grids

Clair Barnes and Wilfrid S. Kendall

## Introduction

The PEML project has organized and collated a substantial quantity of images, and has used this as evidence to support the hypothesis that Anglo-Saxon building construction was based on grid-like planning structures based on fixed modules or *quanta* of measurement. In this appendix we report on the development of some statistical contributions to the debate concerning this hypothesis. In practice the PEML images correspond to data arising in a wide variety of different forms (post-holes, actual walls, indications of linear structure arising from hedge and road lines, etc). Ideally one would wish accurately to reduce each image to a measurement-based dataset that could be analysed statistically to bear on the historical debate. In practice it does not seem feasible to produce a single automatic method which can be applied uniformly to all such images; even the initial chore of cleaning up an image (removing extraneous material such as legends and physical features which do not bear on the planning hypothesis) typically presents a separate and demanding challenge for each different image. Moreover care must be taken, even in the relatively straightforward cases of clearly defined ground-plans (for example for large ecclesiastical buildings of the period), to consider exactly what measurements might be relevant. In this appendix we report on pilot statistical analyses concerning three different situations. These establish not only the presence of underlying structure (which indeed is often visually obvious), but also provide an account of the numerical evidence supporting the deduction that such structure is present. Moreover this approach allows us to map out the range of variation of possible structures suggested by the visual evidence. We contend that statistical methodology thus contributes to the larger historical debate and provides useful input to the wide and varied range of evidence that has to be debated.

Before turning to actual statistical analyses, we note that useful statistics requires a sufficient supply of data (and preferably the data should be in reasonably homogeneous form). The following table illustrates this point by comparing sizes of previous datasets used in comparable investigations, drawing in part from the useful survey of past applications of the quantogram technique to be found in the excellent MSc dissertation of Cox (2009).

| Reference | Dataset | Number of measurements |
|---|---|---:|
| D. G. Kendall (1974) | Diameters of megalithic stone circles in Scotland, England & Wales ("primary" dataset $SEW_2$) | 169 |
| Hewson (1980) | Ashanti goldweights (geometric *versus* figurative design) | 1208+1651 |
| Pakkanen (2004a) | Measurements taken from Temple of Zeus at Stratos (Tables 1, 2) | 17+20 |
| Pakkanen (2004b) | Measurements taken from Toumba building at Lefkandi (column 1 of Table 1) | 27 |
| Cox (2009) | Measurements taken from Sanctuary of Great Gods, Samothrace (Table 5) | 113 |
| W. S. Kendall (2013) | Ground plans of large Anglo-Saxon ecclesiastical buildings | 79 |

Datasets lead to clearer conclusions if they are composed of larger numbers of measurements, though the measurements do need come from sources that are not too inhomogeneous. As can be seen from the table, the dataset discussed in W. S. Kendall (2013) is relatively small (though large enough that one can still detect a clear signal of an underlying quantum). In the next section we revisit the analysis of this dataset, and supplement it with a separate analysis of 110 measurements drawn from the tables of Huggins et al (1982) concerning smaller Anglo-Saxon buildings.

# 1. Anglo-Saxon perches

W. S. Kendall (2013) provides a preliminary analysis of measurements taken from five ground plans of large Anglo-Saxon ecclesiastical buildings. D.G. Kendall's (1974) quantogram technique was modified and applied to determine whether there was evidence for a module or quantum $q$ of measurement underlying the ground plans, in the sense that appropriately chosen measurements $X_i$ (for $i$ running from 1 to $N$) are approximately divisible by $q$. As noted in these references, the quantogram graphs the formula

$$\sqrt{\frac{2}{N}} \sum_i \cos\left(\frac{2\pi X_i}{q}\right)$$

as a function of the frequency $\omega = 1/q$ corresponding to the possible quantum $q$. High peaks of the graph in suitable regions can be interpreted as evidence that $q$ might be a quantum. "High" here is measured in comparison with matched simulations, using random perturbations of the data $X_i$ to generate simulated comparison boundaries.

The first step in applying this technique is to determine the measurements which will constitute the data $X_i$. The Kendall (2013) analysis was based on 79 measurements taken from ground plans, obtained from 13 lines of measurement derived from faces of walls taken from plans of 5 sites (Canterbury SS Peter and Paul, Canterbury St Pancras, Hexham, Escomb, Brixworth). The actual data $X_i$ were derived from all possible differences of measurements in each of the 13 measurement lines. The resulting quantogram led to an estimate of $q = 4.75$m (hence frequency $\omega = 0.21$m$^{-1}$) with a range of error given by $\pm 0.26$m.

An alternative approach, espoused by Huggins (1991), is to use not the faces of walls but the centre-lines, where a centre-line is defined as the mid-line running between the two measured faces of the wall. A simple graph suggests this is probably a better choice: compare the following two plots of measurements for each measurement line. Figure 1 presents measurements of inside and outside faces; Figure 2 presents centre-line averages (noting that measurements have been duplicated where they correspond not to inside and outside wall faces but to boundaries of large objects). Individual measurements are more clearly aligned across separate measurement lines in Figure 2, so for our purposes the centre-line measurements in Figure 2 exhibit a clearer picture and are therefore to be preferred. The averaging process effectively halves the number of distinct measurements, but eliminates non-quantum effects arising from half-widths of walls, if indeed the underlying designs are based on centre-lines.

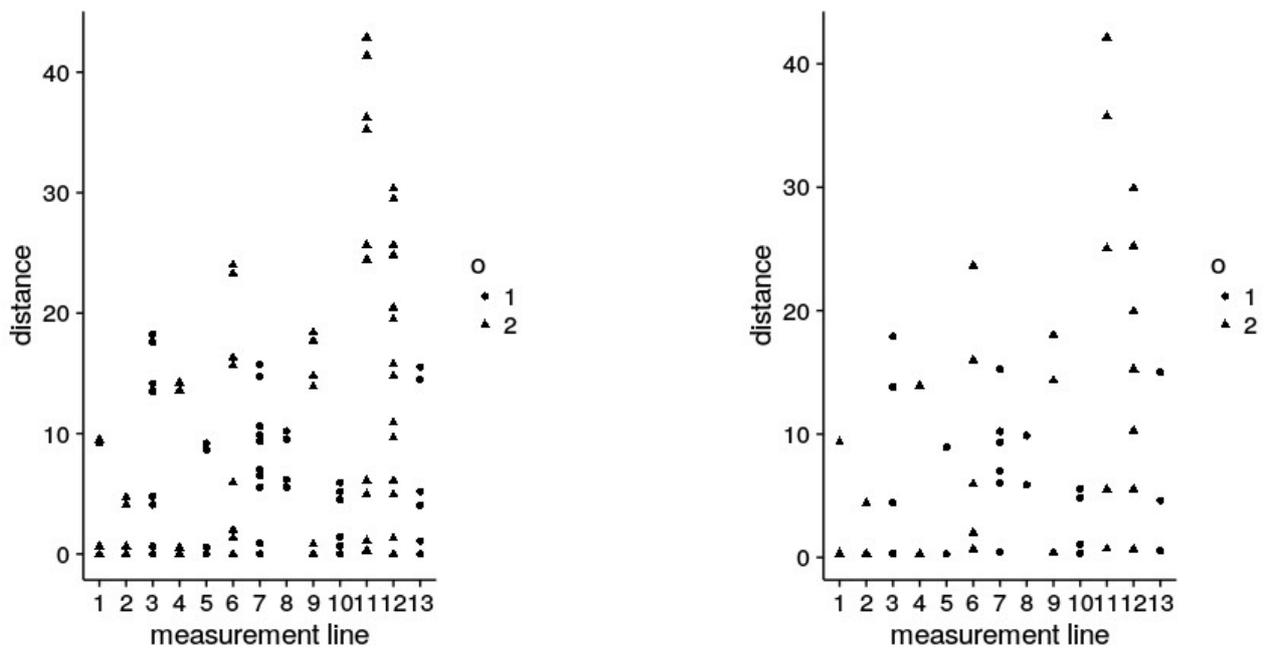

*Figures 1 and 2. Distances in metres based on measurements of wall faces. Shapes indicate wall orientation (E-W or N-S). Figure 1 presents locations of faces of walls; Figure 2 presents centre-lines of walls (duplicating measurements corresponding to internal objects).*

The resulting quantogram, with simulated comparison boundary, is presented in Figure 3. The comparison boundary is obtained by generating 499 simulated quantograms using suitable random perturbations of the data, and then graphing the 5th highest quantogram height at each frequency. This allows us to assess quantogram peaks: a peak that rises well above the simulated comparison boundary is evidence for a genuine underlying quantum. (Determination of the comparison boundary as the 5th highest level of 550=499+1 quantogram levels allows one to interpret this in terms of a statistical test at the 1% level of significance.) Realistic quantum values are expected to lie in the region of 3m to 6.5m (this corresponds to a frequency range of between 0.15m$^{-1}$ and 0.33m$^{-1}$), and in this region the quantogram provides a single well-defined peak at a frequency corresponding to a quantum length of 4.82m, rising clearly above the simulated comparison boundary. Further statistical analysis suggests that a suitable range of error is ±0.26m. Thus this analysis is compatible with the Kendall (2013) analysis (based on facing-wall measurements rather than centre-lines), resulting in a quantum length of 4.75m with error range of ±0.26m.

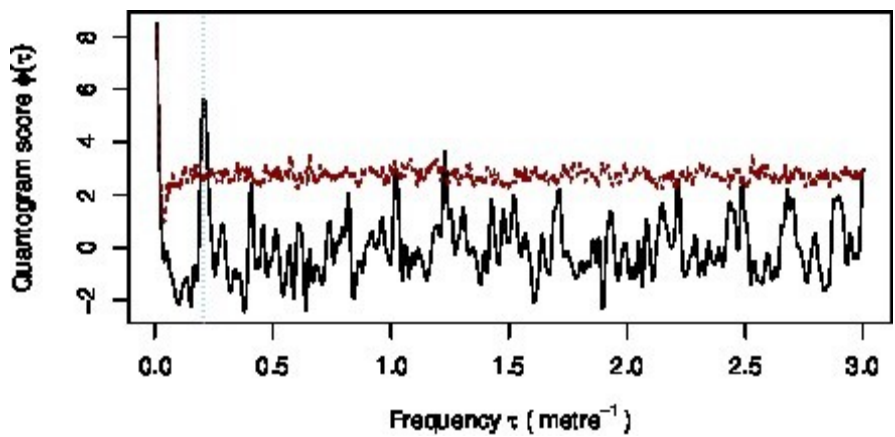

*Figure 3: Quantogram (solid graph) with simulated comparison boundary (irregular dotted graph) for large Anglo-Saxon ecclesiastical buildings (based on all possible pairs of centre-lines).*

Huggins et al (1982) provide a similar dataset relating to smaller buildings: four separate tables containing a total of 110 measurements. These measurements mostly arise as width and depth of separate buildings, rather than as consecutive measurements along selected measurement lines. The resulting quantogram, with simulated comparison boundary, is presented in Figure 4. There are two major peaks, both falling outside the region of interest. However the dominant peak is located at a frequency $\omega = 0.59 m^{-1}$, which corresponds to a quantum level of 1.68m. This is close to one third of the quantum level of 5.05m proposed by Huggins (1991). Bearing in mind that the buildings of this dataset are of smaller size, and that Huggins et al (1982) suggest that building designs could be based on ratios such as 2:3, it is not unreasonable to take this quantogram analysis as lending support to a quantum of 5.05m, Further statistical analysis suggests that a suitable range of error is ±0.25m. Encouragingly, the intervals 4.82±0.26 and 5.05±0.25 overlap.

Evidently it is feasible to produce statistical evaluations of the existence of a quantum or module of measurement, given sufficient data in a suitable form.

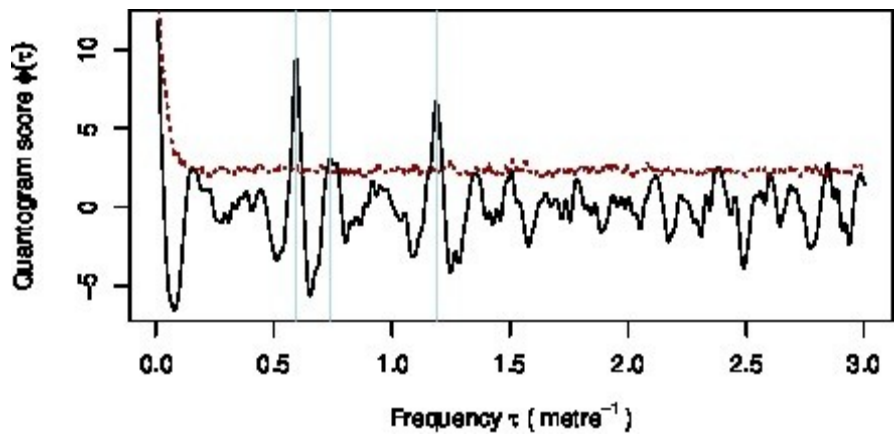

*Figure 4: Quantogram (solid graph) with simulated comparison boundary (irregular dotted graph) for measurements taken from Huggins et al (1982), based on width and depth of separate buildings.*

## 2. Post-holes and Perpendicularity

Much of the PEML data available for analysis does not come in the form of well-defined measurements of distance, but arises indirectly from images of various ground-plans. One seeks to use these images to infer latent linear structure, so obtaining measurements and thus evidence for or against the presence of underlying structure of design. The simplest question of this kind concerns whether or not there is evidence for an underlying design involving organization along perpendicular lines. A good example is provided by the map of Genlis presented in Figure 5.

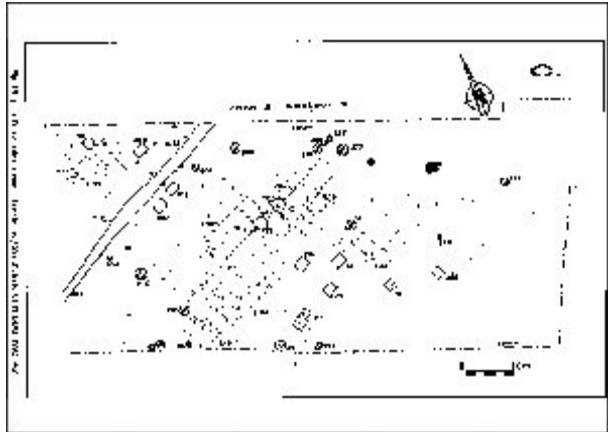

*Figure 5: Genlis map in original form.*

Here interest lies in the information supplied by the post-holes recorded on the map. Visually it appears that many of the post-holes are lined up in a grid-like pattern. The challenge is to connect this visual impression to a statistical and quantitative approach.

The first step is to "data-clean" the image by removing components not related to the post-holes, and then to derive relevant information from the point pattern given by post-hole locations. It is evident that the Genlis image will provide relatively few measurements of distance between walls, and therefore we should not expect this image on its own to supply useful information about quanta. Instead we focus on the question of identifying the walls themselves and then testing whether these walls are organized along perpendicular lines.

Barnes (2015) developed largely automatic methods of image analysis to clean up the Genlis image. These methods involved identification of a list of clumps of connected black pixels, and then removal of those clumps clearly deviating from what one would expect of post-hole images (in other words, removing clumps which were too long and thin or which otherwise deviated from circularity, removing clumps which lay on lines determined by long thin clumps, and removing very isolated clumps). The remaining clumps could be represented as points in a point pattern (Figure 6), and were deemed to represent post-holes.

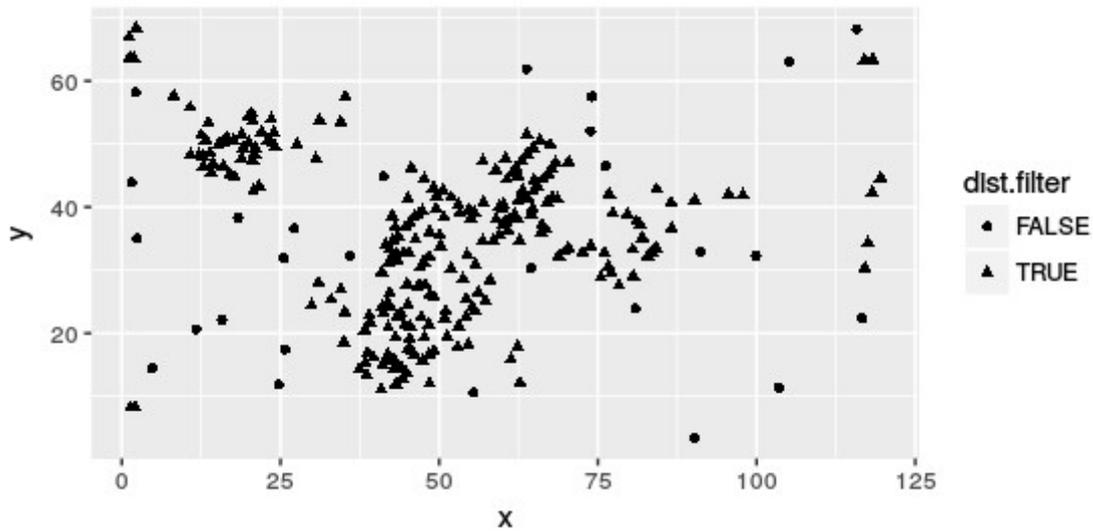

*Figure 6: Post-holes identified from Genlis image after cleaning. Shape indicates whether post-hole is rejected as being too distant from other post-holes (round = rejected; triangle = accepted).*

To each such point was assigned a direction, being the direction towards its nearest neighbour. It was then possible to use these directions to assess evidence for latent perpendicular structure. The initial step of the procedure is to reduce the angles (measured in degrees) modulo 90 degrees so as to produce grid orientations lying in the range 0-90 degrees. In the presence of perpendicular structure, a histogram of the resulting values should then be strongly peaked.

The resulting histogram is presented in Figure 7. It is indeed strongly peaked, though this is supplemented by a uniform spread of some grid orientations across the whole range.

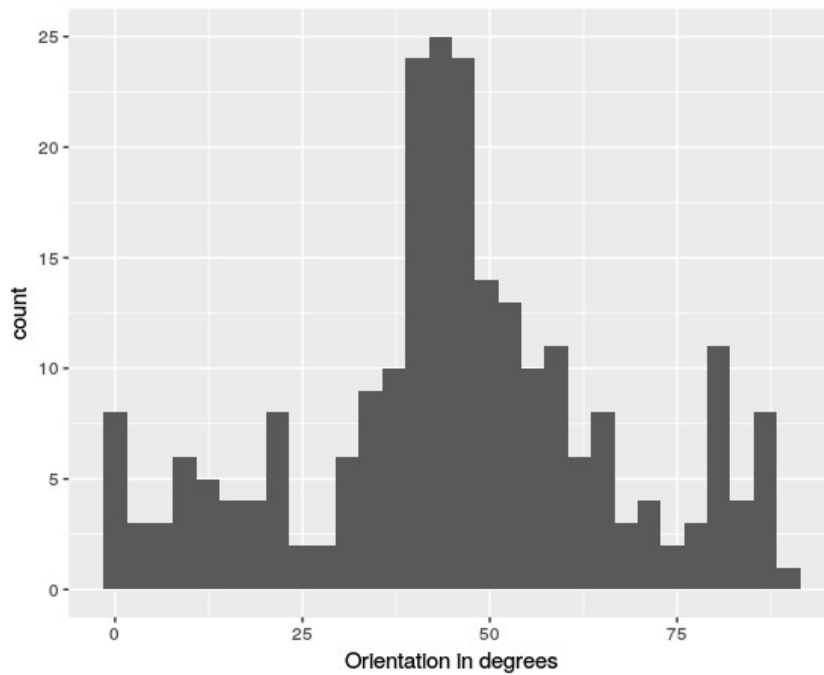

*Figure 7: Histogram of grid orientations for Genlis nearest-neighbour directions..*

It is possible to assess this statistically, by estimating the extent to which this collection of grid orientations is obtained by a mixture of (a) a certain proportion which are spread uniformly across the range (one might suppose these to correspond to rather random post-holes, not aligned to any underlying perpendicular structure) and (b) the remainder being distributed according to some convenient peaked distribution. In this case the convenient peaked distribution turns out to be the von Mises distribution, introduced by Richard von Mises (1918) to assess the fundamental issue in physics of whether atomic weights were multiples of a given unit. (In fact the von Mises distribution also plays a crucial theoretical rôle in justifying the quantogram formula in Section 1, and will again be central to the issue of fitting grids in Section 3. Its rôle in the study of directions and modulus is as fundamental as that of the Normal distribution to the study of location.) Having obtained such a mixture, it is necessary to check that the peak genuinely represents perpendicularity (nearest-neighbour directions clustering in all four directions reducing to the peak modulo 90 degrees), rather than collinearity (nearest-neighbour directions clustering in only two opposite directions). Individual post-hole locations can then be judged to be produced by gridding, or not, depending on a likelihood calculation constructed using the uniform and von Mises distributions estimated for the mixture. Supplemented by a calculation using a spatial clustering algorithm, the result is displayed in Figure 8: post-hole locations are divided into two potentially grid-like and spatially distinct structures together with a substantial number of locations which are not assigned to either structure.

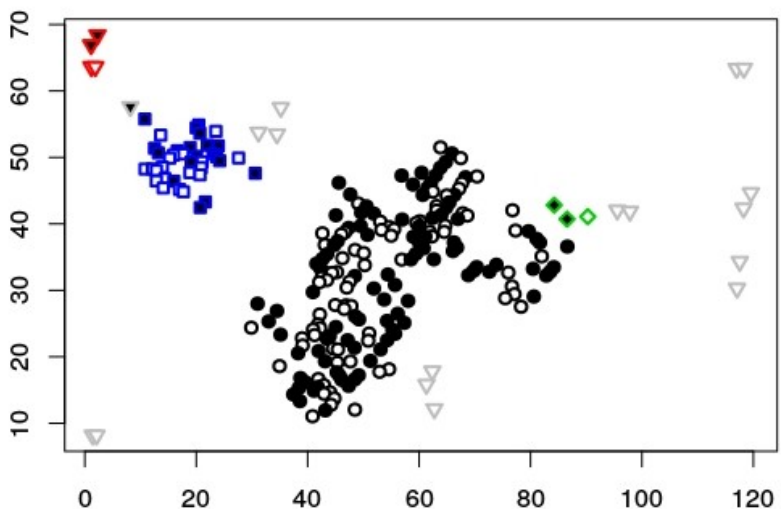

*Figure 8: Black dots indicate post-holes which appear to be part of aligned grid structure. Shapes encode spatial clustering.*

Note that this analysis depends on there being a single underlying grid orientation holding across the image. The task of dealing with an image that contains two overlapping grids would be much more demanding. We have some ideas on how to approach such a task; however we have not yet carried these ideas through to the point of being able to evaluate their practicability. However in some instances it is already possible to make progress, as is illustrated in the next section.

## 3. Post-holes and Grids

We now seek to address the more ambitious challenge of deriving estimates of both quantum length and grid structure from a single image. We consider an image of Brandon (Figure 9), paying particular attention to the post-hole locations.

*Figure 9: Brandon map. Manually identified corner post-holes are indicated here as enlarged dots.*

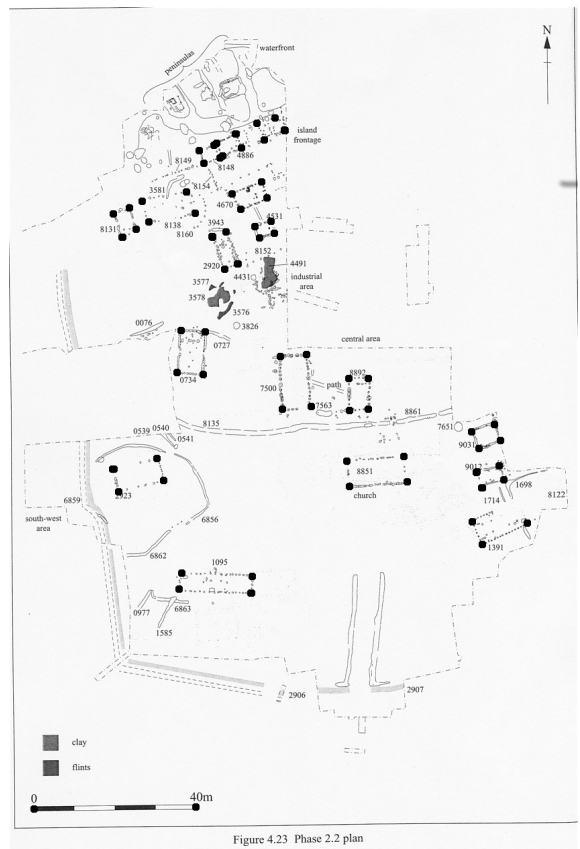

Figure 4.23 Phase 2.2 plan

Inspection of the image reveals some issues. There are many post-holes, typically organized into segments which are nearly straight, but nevertheless can possess some systematic curvature. Rather than devise a statistical technique to deal with this curvature, it was decided to identify corners of buildings manually where possible, and to work with the resulting corner post-hole points. In Figure 9 the corners are highlighted as large black dots: note that in one case the fourth corner of the building is missing.

It then becomes a routine computational task to read in the corner post-hole coordinates and to group the coordinates (manually again) according to the buildings to which the corresponding corner post-holes belong.

The next step is statistical, namely to undertake a quantogram analysis based on individual building widths and depths after the manner of the investigation of the Huggins et al (1982) data in Section 1 above. This analysis showed no evidence for a significant peak at all (not entirely a surprise: the 17 Brandon buildings are mostly rather small). In order to find structure, it is necessary to consider the buildings in relation to each other. However, as is clear from Figure 9, there appear to be two distinct grid patterns. To have any chance of success, it is necessary to determine which buildings correspond to which grids.

The methods of Section 2 suggest an effective approach. Consider the directions given by each of the building edges. Reducing angles modulo 90 degrees as before, we obtain the following histogram of grid orientations (Figure 10: note that the data here are rotated through 45 degrees to avoid splitting a cluster of orientations between the right- and left-hand sides of the histogram).

*Figure 10: Histogram of grid orientations for building sides: there is clear evidence of two different overlapping clusters.*

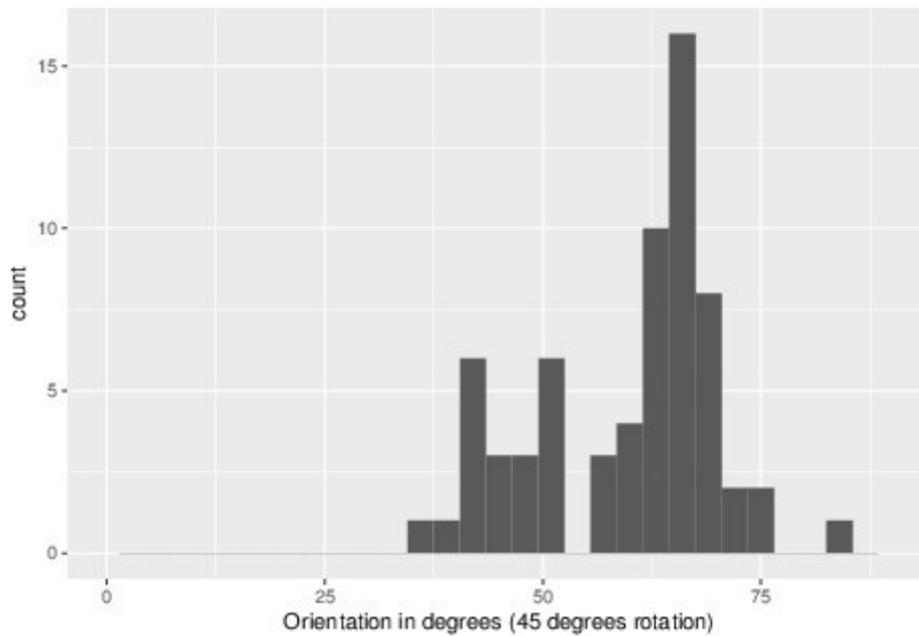

The histogram suggests the existence of two overlapping clusters. Again it is possible to assess this statistically, this time by estimating the extent to which this collection of grid orientations is obtained by a mixture of two different von Mises distributions. The best-fitting mixture provides a reasonable statistical fit, making it possible to assign each of the buildings to one of the two grid orientations represented by the mode of the von Mises distributions, depending on whether all edges of the building in question are assigned to one or to the other grid orientation (Figure 11). Note that one of the buildings is left unassigned as its edges were not all assigned to the same grid orientation; this building is omitted from the subsequent analysis.

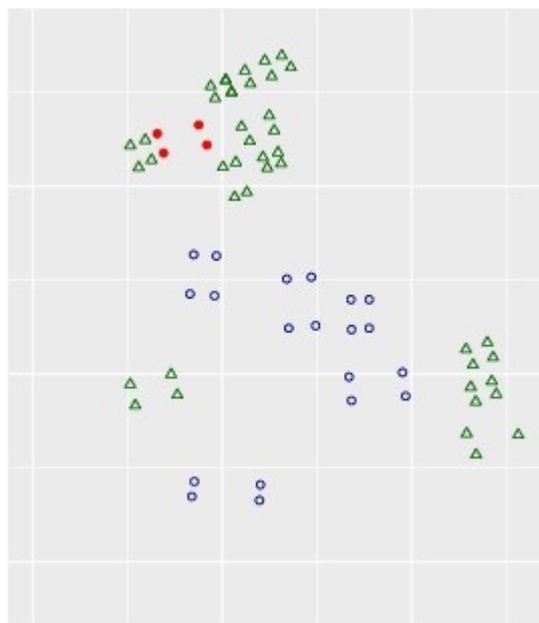

*Figure 11: The two different clusters of buildings (corners indicated respectively by open circles and triangles) and the unassigned building (corners indicated by solid circles).*

For each grid orientation one can now calculate two perpendicular axes of measurement; we will refer to these as x and y axes. Effectively each post-hole belonging to each of these buildings provides an x coordinate and a y coordinate determined by the corresponding grid system. The quantogram technique can now be applied to the compendium dataset obtained as the union of the four sets of all possible differences between coordinates; one set of all possible differences being constructed for each of the four measurement axes arising from the two grids. The resulting quantogram, with simulated comparison boundary, is presented in Figure 12.

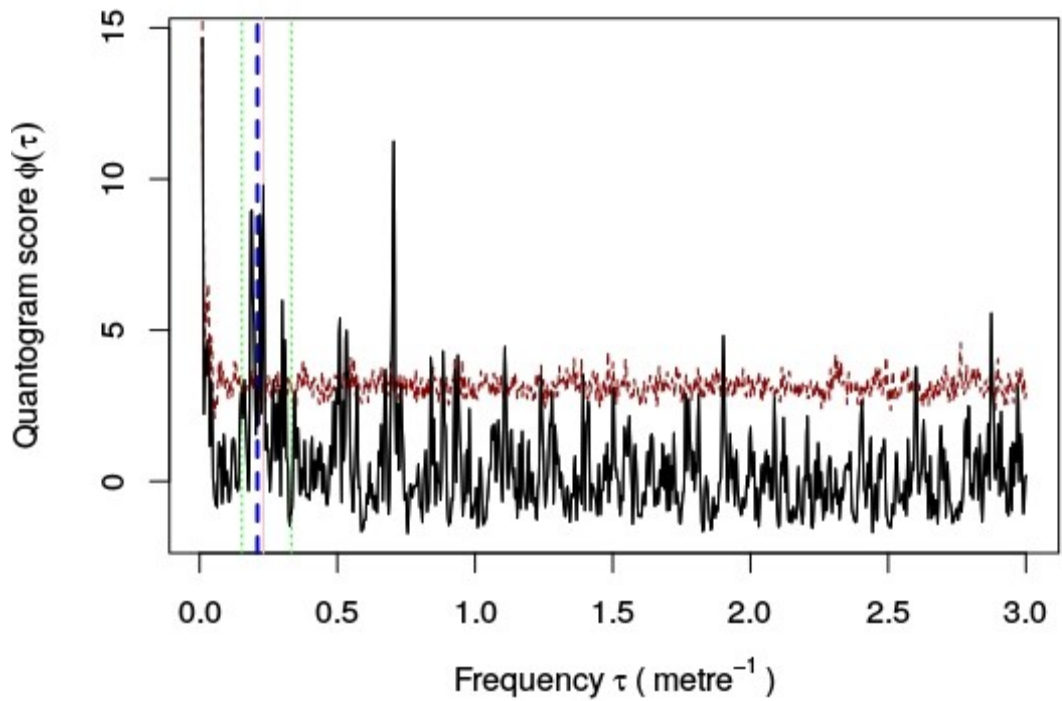

*Figure 12: Quantogram (solid graph) with simulation comparison boundary (irregular dotted graph) using compendium of all distances from Brandon data. The solid vertical line indicates the frequency yielding highest level within region of interest delimited by thin vertical dotted lines). The dashed vertical line indicates the frequency corresponding to 4.75m.*

Here the dominant peak in the region of interest corresponds to a quantum of 4.32m. However there are a number of neighbouring peaks, not quite so high. A vertical dashed line indicates a smaller peak located at the frequency corresponding to the quantum of 4.75m used elsewhere in this monograph, and it is reasonable to take the view that the quantogram provides evidence for a quantum in this range but without distinguishing (for example) between these two quanta.

There is a very large peak outside of the region of interest, located at one third of the quantum of 4.32m, and thus possibly related to systematic construction based on a quantum 4.32m.

Bearing in mind the number of closely neighbouring peaks, and the chain of statistical procedures required to reach our conclusions, we omit the analysis of range of error and instead simply note that the estimates of quantum size based on this quantogram should be viewed as indicative only.

It should be noted that quantograms were also constructed for each of the four separate measurement lines. For example, taken separately the smaller cluster provides rather weak evidence for a larger quantum of 6.38m. However the corresponding peak disappears in the compendium analysis of both clusters taken together. Subsequent calculations are based on use of the quantum obtained by the compendium analysis, and compared with corresponding calculations using the quantum of 4.75m.

Given a quantum and a grid orientation, one has of course to select from a range of possible grids depending on choice of x and y offset. However here again the von Mises distribution finds employment: for each cluster we can compute the x and y offsets of each post-hole from the grid, fit von Mises distributions to the cluster population of x offsets and separately the population of y offsets, and use the modes of the two von Mises distributions to determine the offsets for the actual grid to be fitted to the cluster. (In fact the theory of the von Mises distribution shows that this procedure, shorn of its statistical context, corresponds to a natural circular averaging process.) Figures 13 and 14 respectively illustrate the grids fitted on the basis of a quantum of 4.32m and a quantum of 4.75m. Visually the difference is not large. On closer inspection, the quantum of 4.32m does appear to provide a better fit. In both cases, the fit to the larger and more spread-out cluster (indicated by open circles, and with grid more at a slant) appears to be better than the fit to the smaller and more compact cluster. Note that in both cases the fit might be improved by excluding a couple of less-well-aligned buildings.

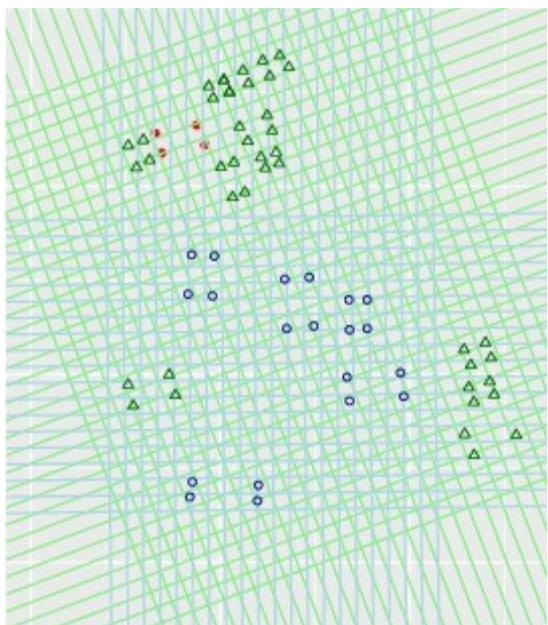 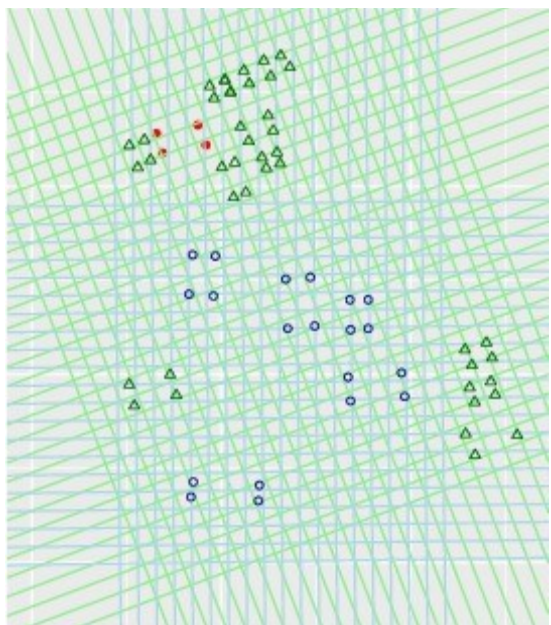

*Figure 13 (left): Fitted grids for the plot of corners using quantum (4.32m) obtained from Brandon data: cluster 1 uses triangles and cluster 2 uses open circles.*

*Figure 14 (right): Fitted grids for the plot of corners using previous quantum (4.75m) obtained from large Anglo-Saxon ecclesiastical buildings: cluster 1 uses triangles and cluster 2 uses open circles.*

## Conclusion

The work above demonstrates that it is possible to produce estimates of quanta and fitted grids for suitable images and data sources, and that the resulting fit can be explored statistically. These statistical treatments supplement, but do not replace, the informed manual analyses discussed in the remainder of this monograph. As noted above, the variety of images means that each separate image requires its own data-cleaning approach, and will typically present further challenges arising from the different sources of data present in the image. We have shown how relevant information can be extracted statistically from some well-behaved images and data-sources: this contributes to the debate and particularly emphasizes the extent to which estimates (whether of quanta or of grids) should be accompanied by ranges of accuracy.

The statistical approach given here is still only a preliminary approach: the pervasive presence of the von Mises distribution suggests the possibility of assimilating the present layered statistical approach (determine grid orientations, if satisfactory then fit quanta, if satisfactory then choose grids) into a more unified treatment. However a more pressing question is, how can one deal with grids which are not as well separated as in the Brandon example? This is a substantial question, and it may well be that overlaid grids simply present too great a challenge. However we hope to explore the issue in future work.

## Acknowledgements

This work was partly supported by EPSRC Research Grant EP/K031066/1. We thank John Blair for supplying the measurement data for ground plans of large Anglo-Saxon ecclesiastical buildings as discussed in Section 1, the Genlis image discussed in Section 2, and the Brandon image discussed in Section 3. All data created during this research, together with associated R scripts expressed in *Rmarkdown* format (http://rmarkdown.rstudio.com/), will be openly available from the University of Warwick's institutional repository (http://wrap.warwick.ac.uk/75491). The dataset of smaller Anglo-Saxon buildings, also discussed in Section 1, is available in Tables 1, 3, 4, 5 of Huggins et al (1982).